# Influence of a transverse static magnetic field on the magnetic hyperthermia properties and high-frequency hysteresis loops of ferromagnetic FeCo nanoparticles.

B. Mehdaoui, J. Carrey*, M. Stadler, A. Cornejo, C. Nayral, F. Delpech, B. Chaudret and M. Respaud

Université de Toulouse; INSA; UPS; LPCNO (Laboratoire de Physique et Chimie des Nano-Objets), 135 avenue de Rangueil, F-31077 Toulouse, France and
CNRS; UMR 5215; LPCNO, F-31077 Toulouse, France

**Abstract:**
The influence of a transverse static magnetic field on the magnetic hyperthermia properties is studied on a system of large-losses ferromagnetic FeCo nanoparticles. The simultaneous measurement of the high-frequency hysteresis loops and of the temperature rise provides an interesting insight into the losses and heating mechanisms. A static magnetic field of only 40 mT is enough to cancel the heating properties of the nanoparticles, a result reproduced using numerical simulations of hysteresis loops. These results cast doubt on the possibility to perform someday magnetic hyperthermia inside a magnetic resonance imaging setup.

**Main Text:**

Magnetic hyperthermia (MH) is a promising cancer treatment technique based on the fact that magnetic nanoparticles (MNPs) placed in an alternating magnetic field release locally heat [1, 2, 3]. Its efficiency in combination with radiotherapy has been recently demonstrated in the treatment of glioblastome multiforme [4]. On the long term, performing magnetic hyperthermia in a magnetic resonance imaging (MRI) setup could have several advantages: i) The production of the magnetic field for MH could be produced by the same hardware as for the MRI. ii) The measurement of the local temperature during hyperthermia treatment using MRI thermometry could allow for a better control of the treatment, by avoiding overheating effect in the safe zones. iii) MRI can quantify of the local concentration of nanoparticles trapped inside the tumor, which is essential to define the treatment parameters.

However, the presence of a static magnetic field in MRI set-ups should decrease the specific absorption rate (SAR) of the MNPs, as pointed out by a few theoretical works [5, 6, 7, 8]. This explains why the development of a combined system should probably be based on a low-field MRI setup working a static field of 0.1 or 0.2 T only [5]. So far, no experimental study of the influence of a static magnetic field on the MH properties of MNPs has been reported. In the present article, we report such an investigation for a model system of ferromagnetic FeCo nanoparticles. In a first study, we demonstrated that these NPs display large losses and a behavior typical of the Stoner-Wohlfarth regime [9]. To go deeply into the mechanisms involved, the influence of the static magnetic field on MH properties is studied by combining temperature measurements and high-frequency hysteresis loop measurements. Our results, which are consistent with numerical calculations of the hysteresis loops, show that a small magnetic field is sufficient to cancel the heating properties of the MNPs.

The FeCo nanoparticles studied in the present article are shown in Fig. 1(a). Their synthesis method has been described elsewhere [9, 10]. Their mean diameter distribution is bimodal with two peaks centered at 5.5±1.1 nm and at 12.8±2.0 nm, as measured by transmission electron microscopy (TEM). The mass concentration of metal in the powder is 89.6±5.0 % and the ratio Fe/Co is 1.46±0.30, as determined by chemical analysis. The saturation magnetization $\sigma_S$ of the nanoparticles using this synthesis method is 140±8 Am$^2$.kg$^{-1}$, well below the bulk value (240 Am$^2$kg$^{-1}$), due to an imperfect alloying of Fe and Co atoms, and to the high carbon content [10]. For the MH studies, the investigated sample is a sealed vessel containing 13.0 mg of powder, 32 mg of hexadecylamine, and 388 mg of tetrahydrofuran.

The MH properties have been characterised through the simultaneous measurement of the temperature rise and of magnetic hysteresis loops on a single setup schematized in Fig. 1(b). The alternating magnetic field $\mu_0 H_{AC}$ is generated by an air-cooled Litz wire coil. Inside this coil a system comprising a calorimeter and two contrariwise-wounded compensated pick-up coils connected in series is mounted. The two pick-up coils have the same section surface $S_{coil}$ and number of turns $n$. The whole setup is placed inside a standard electromagnet which produces the perpendicular static magnetic field $\mu_0 H_{DC}$. The amplitude of the alternating magnetic field is obtained using $\mu_0 H_{AC} = \dfrac{\int e_1 dt}{nS_{coil}}$, where $e_1$ the voltage at the terminals of the empty coil. The magnetization per unit mass $\sigma_S$ of the nanoparticles is obtained using $\sigma_S = \dfrac{\int e_2 dt}{\mu_0 n \rho S_{vessel} C}$, where $S_{vessel}$ is the surface of a section of the vessel containing the colloidal solution, $C$ the volume concentration of the sample, $\rho$ the MNP density and $e_2$ the voltage at the terminal of the two coils in series. For temperature measurements, 1.5 ml of water is added around the vessel to ensure its thermalisation. To measure the specific absorption rate (SAR), the temperature rise of the water is measured after the magnetic field has been applied during 100 s. The raw values are multiplied by a factor 1.20 to take into account the losses of the calorimeter, which were previously determined. The specific losses $A$ of the samples (expressed in mJ/g) are then estimated by two methods: i) from magnetic measurements, by integrating the hysteresis loops, or ii) from temperature measurements, using the equation $SAR = Af$, where $f$ is the frequency of the alternating magnetic field.

The hysteresis loops measured at $f$ = 56 kHz for various values of $\mu_0 H_{AC}$ in the absence of any transverse static magnetic field are shown in Fig. 2(a). These hysteresis loops are nearly square and saturated, which indicates that the NPs have a ferromagnetic behaviour and an anisotropy axis oriented along the direction of the field [3]. Under the application of the magnetic field, the nanoparticles are expected to orient their individual anisotropy axis along the direction of the field and/or form chains of nanoparticles. Each phenomena or the combination of both are likely to explain the observed anisotropy axis of the sample. The magnetization for $\mu_0 H_{AC}$ = 23 mT is $\sigma_S$ = 33 ± 5 Am$^2$.kg$^{-1}$ below the one extracted from quasi-static SQUID measurements (140±8 Am$^2$.kg$^{-1}$). This discrepancy probably comes from the small particles seen in TEM, which are superparamagnetic and far from being saturated for such a small magnetic field. Fig. 2(b) shows the specific losses $A$ deduced from the integration of the hysteresis loops shown in Fig. 2(a) and the ones deduced from temperature measurements. The agreement between both validates the quality of our experimental setup and the method of calculation.

Fig. 3(a) shows the influence of a transverse static magnetic field on the high-frequency hysteresis loops when an alternating field $\mu_0 H_{AC} = 23$ mT is applied. A small static magnetic field of 8 mT is enough to significantly reduce the hysteresis loop area, its squareness and the overall susceptibility of the sample. When $\mu_0 H_{DC} = 34$ mT, the hysteresis loop appears almost fully closed and reversible. For $\mu_0 H_{DC} \approx 1$ T, the susceptibility is almost null. In Fig. 3b, the evolution of the hysteresis area with $\mu_0 H_{DC}$ deduced from the integration of these hysteresis loops are shown and compared with temperature measurements. Both match well and confirm that the heating of the nanoparticles is fully cancelled when $\mu_0 H_{DC}$ exceeds 40 mT.

To check if this behaviour is surprising or can be theoretically expected, we have performed numerical simulations of the hysteresis loops. These simulations have been described in details in Ref. [3] and fully tested; they are able to calculate the hysteresis loop area $A$ for an assembly of magnetically independent spherical uniaxial NPs with their anisotropy axis randomly distributed in space or aligned with the magnetic field. We have already shown that such simulations can reproduce several features observed experimentally in MH [11]. The model has been modified in order to take into account a transverse static magnetic field. Here, we assume that the NP anisotropy axes are aligned with the alternating magnetic field and we fixed the following set of parameters: the Néel-Brown frequency factor $\tau_0 = 5 \times 10^{-10}$ s, $\sigma_S = 140$ Am$^2$kg$^{-1}$, the temperature $T = 300$ K and $f = 56$ kHz. We have set the effective anisotropy of the NPs $K_{eff}$ and their radius $r$ as free parameters. In order to neglect the presence of the small superparamagnetic nanoparticles, which do not contribute to the SAR and thus artificially decrease the experimental SAR value, the fit was performed on the normalized experimental curves only. To fit the experimental results, we have tried to reproduce the shape of the experimental hysteresis loops and the evolution of their areas versus the applied ac and dc applied magnetic fields $A(\mu_0 H_{AC})$ and $A(\mu_0 H_{DC})$. The parameters found for the fit are in the range $K_{eff} = 2.1$ to $4.0 \times 10^4$ J.m$^{-3}$ and $r = 6$ to 9 nm, in agreement with the TEM results and with our previous estimation of the anisotropy [9]. The numerical results for $K_{eff} = 2.8 \times 10^4$ J.m$^{-3}$ and $r = 7.5$ nm are shown in Figs 3(b)-(d). An important point is that, in the range of $K_{eff}$ and $r$ values permitting acceptable fits of our experimental data, the decrease of the SAR with $\mu_0 H_{DC}$ was always stronger or equal to the one found experimentally.

To go further, we have performed some prospective simulations to check if there is a way to diminish the influence of the transverse magnetic field on the SAR (not shown). From our preliminary calculations, one can identify two possibilities, through the increase of the NP anisotropy and/or the application of very large $\mu_0 H_{AC}$ value. The former solution however leads to a decrease of the maximum SAR [3] and the latter is problematic for technical and/or biomedical reasons. Even if these theoretical and experimental studies should be pursued, it seems that large *SAR*s and insensitivity to static magnetic field are hardly-compatible requirements.

In conclusion, these first experimental results on the influence of a transverse magnetic field on MH properties measured on a model system consisting of ferromagnetic FeCo NPs, shows that a static magnetic field as small as 40 mT is enough to completely cancel their heating properties, which is consistent with numerical simulations. These first results are rather pessimistic on the possibility to perform MH inside an MRI setup.

**Acknowledgements :**

This work was supported by the Fondation InNaBioSanté, the Région Midi-Pyrenées and European Commission for the POCTEFA Interreg project (MET-NANO EFA 17/08). A.C. is grateful to the Spanish Ministerio de Ciencia e Innovacion for a postdoctoral grant. We thank Angélique Gillet for her helpful work and AVAMIP for funding.

**References :**

[*] Electronic mail: julian.carrey@insa-toulouse.fr

**Figures :**

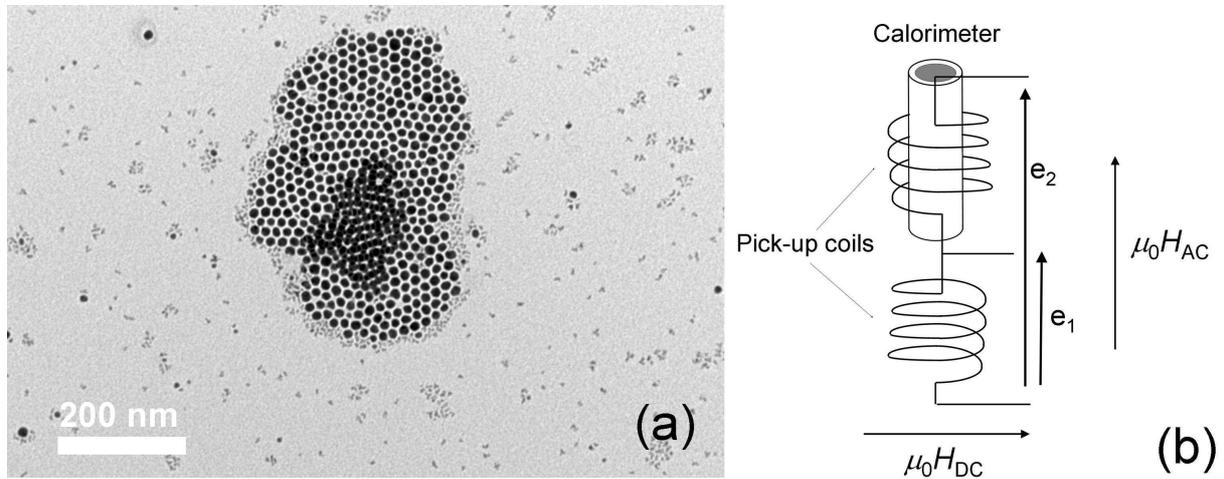

Figure 1: (a) TEM micrograph of the FeCo nanoparticles. (b) Schematic representation of the experimental setup.

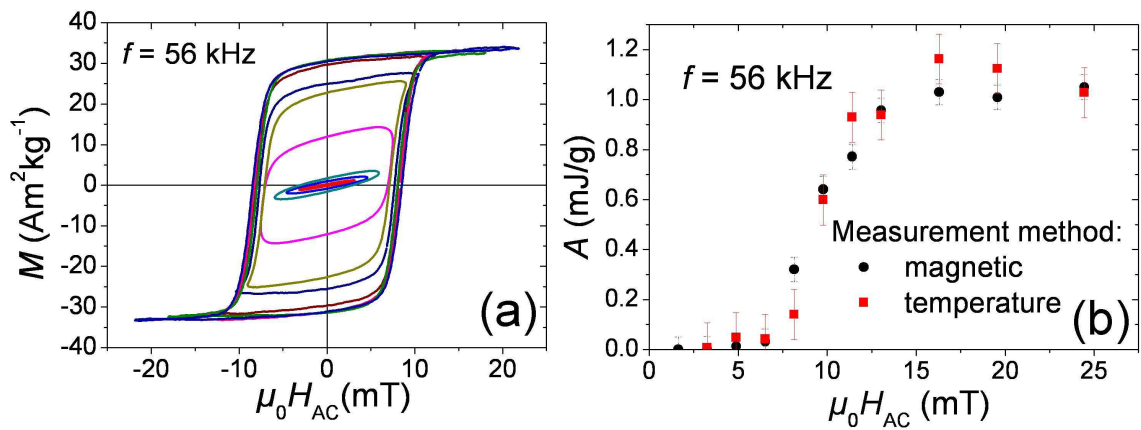

Figure 2 (color online): (a) Hysteresis loops measured for various values of $\mu_0 H_{AC}$. (b) Comparison between the specific losses deduced from the integration of hysteresis loops and from temperature measurements. Note that no static transverse field is applied here.

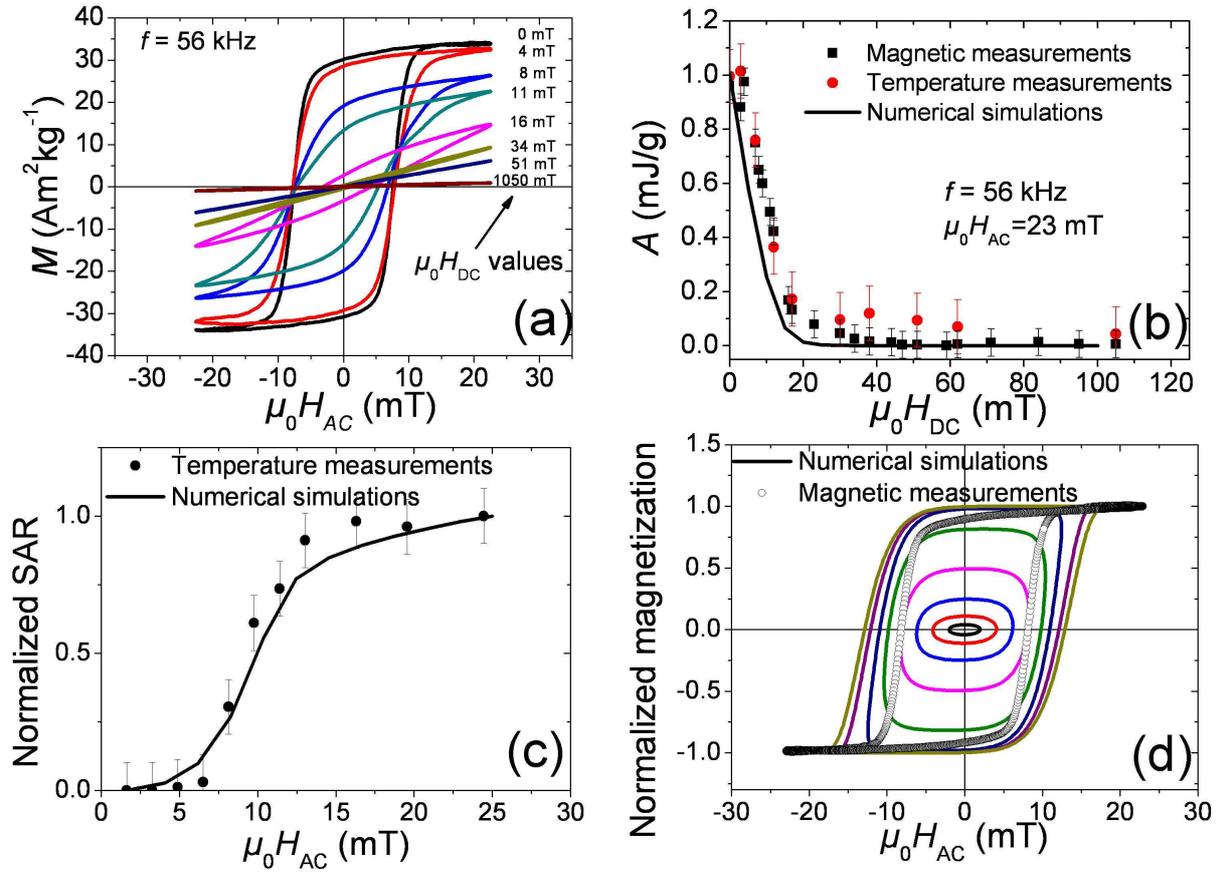

Figure 3 (color online): (a) Hysteresis loop measurements as a function of the static magnetic field $\mu_0 H_{DC}$ for $\mu_0 H_{AC}$=23 mT. (b) Decrease of hysteresis area value with $\mu_0 H_{DC}$ for $\mu_0 H_{AC}$=23 mT. The comparison between magnetic measurements, temperature measurements and numerical simulations is shown. For the numerical simulations, the parameters are $\tau_0 = 5\times10^{-10}$s, $\sigma_S = 140$ Am$^2$kg$^{-1}$, $T = 300$ K, and $f = 56$ kHz, $K_{eff} = 2.8\times10^4$ J.m$^{-3}$ and $r = 7.5$ nm. Numerical simulations results are normalized with respect to the experimental zero field value. (c) Comparison between the magnetic field dependence of normalized SAR extracted from temperature measurements and numerical simulations for $\mu_0 H_{DC} = 0$ mT. (d) Comparison between hysteresis loop measurements and numerical simulations for $\mu_0 H_{DC} = 0$ mT. The numerical simulations are shown for various values of $\mu_0 H_{AC}$ and the experimental curve is the one for $\mu_0 H_{AC}$=23 mT.